\documentclass{article}
\usepackage[preprint]{spconf}
\copyrightnotice{\copyright\ IEEE 2024}
\toappear{To appear in {\it Proc.\ ICASSP2024,
                    April 14-19, 2024, Seoul, Korea}}
\usepackage{spconf,amsmath,graphicx}
\usepackage{bm}
\usepackage{multirow,multicol}
\usepackage{makecell}
\usepackage{bbding}
\usepackage{threeparttable}

\title{{\Large F}ast{\Large I}nject: Injecting Unpaired Text Data into CTC-based ASR training}
\name{Keqi Deng, Philip C. Woodland\thanks{Keqi Deng is funded by the Cambridge Trust. This work has been performed using resources provided by the Cambridge Tier-2 system operated by the University of Cambridge Research Computing Service (www.hpc.cam.ac.uk) funded by EPSRC Tier-2 capital grant EP/T022159/1.}}
\address{Department of Engineering, University of Cambridge, Trumpington St., Cambridge, UK.\\\small{\texttt{\{kd502, pcw\}@eng.cam.ac.uk}}}
\begin{document}
%
\maketitle
\begin{abstract}
Recently, connectionist temporal classification (CTC)-based end-to-end (E2E) automatic speech recognition (ASR) models have achieved impressive results, especially with the development of self-supervised learning. However, E2E ASR models trained on paired speech-text data often suffer from domain shifts from training to testing. To alleviate this issue, this paper proposes a flat-start joint training method, named FastInject, which efficiently injects multi-domain unpaired text data into CTC-based ASR training.
To maintain training efficiency, text units are pre-upsampled,
and their representations are fed into the CTC model along with speech features.
To bridge the modality gap between speech and text, an attention-based modality matching mechanism (AM3) is proposed, which 
retains the E2E flat-start training.
Experiments show that the proposed FastInject gave a 22\% relative WER reduction (WERR) for intra-domain Librispeech-100h data
and 20\% relative WERR on out-of-domain test sets.

\end{abstract}
%
\vspace{-0.1cm}
\section{Introduction}
\label{sec:intro}
\vspace{-0.1cm}
End-to-end (E2E) automatic speech recognition (ASR) directly transcribes speech into text \cite{graves2006connectionist, Graves2012SequenceTW}.
Among E2E models, the connectionist temporal
classification (CTC)-based \cite{graves2006connectionist} model is popular due to its simple structure and fast decoding as a non-autoregressive model \cite{lee2021intermediate}.
With the development of self-supervised pre-training \cite{NEURIPS2020_92d1e1eb, zhang2022speechlm}, CTC performance can surpass pipeline methods \cite{deng2022improving}. However, E2E ASR still suffers from domain shifts \cite{10095860, deng2023label} and even the CTC-based model has been shown to learn an implicit internal LM \cite{kanda2016maximum,das2023mask}, which characterises the training data distribution and degrades generalisation \cite{das2023mask}. This issue can be alleviated by training on large amounts of multi-domain speech-text paired data, however, collecting this hand-transcribed data is expensive compared to text-only data \cite{10095860, sainath2023joist}. Hence, building a robust CTC-based model using text-only data is more efficient.

Training an external language model (LM) to improve the E2E ASR via shallow fusion \cite{chorowski2015attention} is a common solution to utilise text-only data. In addition, to handle the internal LM bias of E2E ASR, a density ratio method \cite{9003790} was proposed to estimate the internal LM score using a source-domain external LM. However, this estimation process complicates the decoding and is not always accurate \cite{tsunoo22_interspeech}. Moreover, applying external LMs to the CTC-based model loses the advantage of fast decoding speed as a non-autoregressive model \cite{deng2022improving}. Therefore, an alternative solution is directly training the CTC-based model with text-only data.


To inject the text into ASR training, 
\cite{Chen2022MAESTROMS} generates alignments on the fly during training to make text representations match the speech feature length, thus enforcing frame-level similarity, but this reduces training efficiency. To avoid this,
\cite{chen2021injecting} no longer explicitly enforces this similarity, but instead relies on a neural transducer-based model to implicitly align them. However, for CTC models with simple structures, enforcing similarity to speech features is still necessary. This paper proposes a flat-start joint training method, FastInject, that injects text data into CTC-based ASR training, while efficiently bridging the modality gap without needing previously obtained alignments. FastInject up-samples text units via random repetition in an offline manner to maintain high training efficiency and inputs text representations
into the CTC model along with speech features. 
Since the Transformer encoder used in CTC learns alignments via self-attention mechanisms, an attention-based modality matching mechanism (AM3) is proposed. AM3 efficiently bridges the modality gap by enforcing the similarity of outputs obtained using speech and text representations respectively as the key-value pairs of dot-product attention.
Hence, since AM3 is highly parallel,
FastInject efficiently learns consistent representations across modalities
without previously obtained alignments, retaining high training efficiency and E2E flat-start training.
Experiments across LibriSpeech-100h \cite{7178964}, GigaSpeech \cite{chen21o_interspeech}, TED-LIUM2 \cite{rousseau-etal-2014-enhancing}, and AESRC2020 \cite{9413386} data show that FastInject greatly boosts the intra and out-of-domain ASR accuracy.

The rest of this paper is organised as follows,
Sec.~2 reviews related work. Sec.~3 describes the proposed methods. Sec.~4 details the experiments and Sec.~5 draws conclusions.

\begin{figure*}[t]
    \centering
    \includegraphics[width=0.96\linewidth]{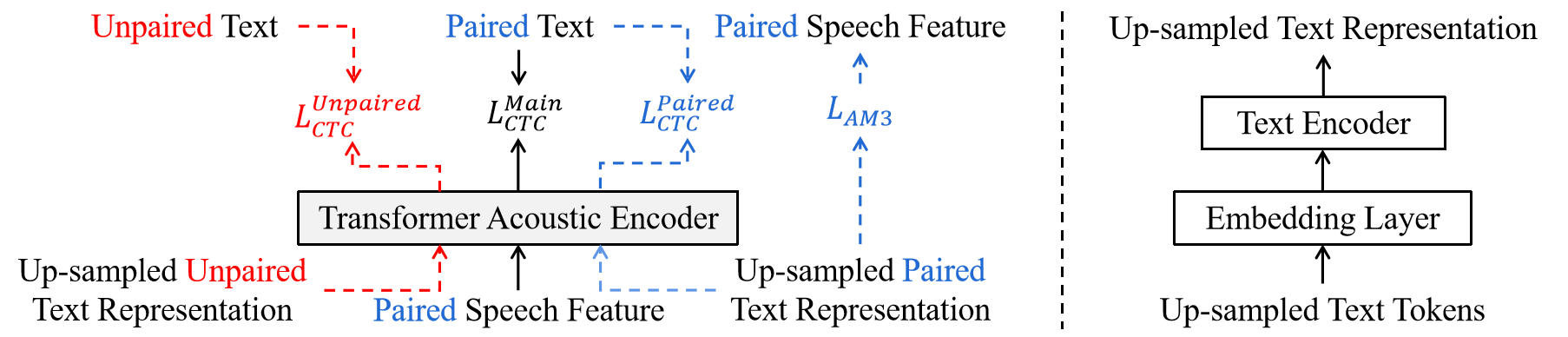}
    \vspace{-0.2cm}
    \caption{Illustration of the proposed FastInject. On the left, red represents unpaired text data and blue denotes paired speech-text data. Dashed lines indicate auxiliary components added during training and dropped during decoding. The right part shows the extraction of the up-sampled text representation. Note the text encoder and the acoustic encoder are trained E2E from scratch. }
    \vspace{-0.16cm}
    \label{fig:arch}
\end{figure*}

\vspace{-0.13cm}
\section{Related Work}
\label{related}
\vspace{-0.12cm}
Injecting text into the self-supervised speech pre-training task
has been widely studied,
including using TTS \cite{chen2021injecting} or mapping speech and text into a shared space \cite{zhang2022speechlm, Ao2021SpeechT5UE, bapna2021slam}.
However, the following fine-tuning could be prone to forgetting the pre-trained task \cite{sainath2023joist,bai2022joint}. To directly inject text-only data into the ASR training stage,  \cite{DBLP:conf/interspeech/WangSW21} designed a two-stage training for the attention-based encoder-decoder (AED) model. In \cite{Chen2022MAESTROMS} up-sampled text representations were generated according to a duration prediction model, which was jointly trained on the alignments generated iteratively by a neural transducer, but this reduces training efficiency. In \cite{sainath2023joist} simpler duration model strategies were explored like random repetition of text representations.
Previous work \cite{sainath2023joist, Chen2022MAESTROMS, DBLP:conf/interspeech/WangSW21, sainath2020attention, peyser23_interspeech} focused on AED \cite{DBLP:conf/interspeech/WangSW21, sainath2020attention} or neural transducer \cite{sainath2023joist, Chen2022MAESTROMS, peyser23_interspeech} models.
These models, unlike CTC, possess a decoder or prediction network that takes text as input, and their more complex structure enhances robustness against mismatch between text and speech representations. For example, \cite{sainath2023joist} still improved ASR results by 
inputting up-sampled text representations into the neural transducer-based structure just like speech features, even without a certain loss function to enforce consistency.
To the best of our knowledge, this is the first work to inject unpaired text data into the CTC-based E2E ASR training without speech synthesis.

\vspace{-0.2cm}
\section{CTC Model with {\large F}ast{\large I}nject}
\label{sec:method}
\vspace{-0.1cm}
This paper proposes FastInject to inject unpaired text data into CTC-based ASR training, as shown in Fig.~\ref{fig:arch}. The motivation is to 
efficiently learn representations from unpaired text data that are consistent with speech features during ASR training.
This paper aims to enhance CTC accuracy while retaining training efficiency and the simple CTC structure.

\vspace{-0.2cm}
\subsection{Unpaired Text Processing}
\vspace{-0.1cm}
To create text representations easily aligned with speech features, prior work \cite{Chen2022MAESTROMS} replicates the text representations on the fly to match the speech duration. 
This is hard to implement in parallel because the number of replications varies depending on the text unit.
To maintain a high training efficiency, this paper up-samples the text units\footnote{This paper uses phone units but also allows for any other units.} by random repetition, which is an offline process and can be regarded as data pre-processing. 
To extract text representations, the up-sampled text units are fed into an embedding layer and text encoder, which are discarded after ASR training.

\vspace{-0.2cm}
\subsection{Attention-based Modality Matching Mechanism}
\label{sec:am3}
\vspace{-0.1cm}
When the text representations have been obtained, the next step involves bridging the gap between text and speech modalities. An attention-based modality matching mechanism (AM3) is proposed that can achieve this with a flat-start.
Suppose the representations obtained from up-sampled paired text units as $\mathbf{P}=(\bm{p_1}, \cdots, \bm{p_L})$, and that obtained from unpaired text units as $\mathbf{U}$, AM3
enforces consistency between $\mathbf{P}$ and paired speech features $\mathbf{S} = (\bm{s_1}, \cdots, \bm{s_T})$  using dot-product attention:
\begin{eqnarray}
    \mathbf{S}^{'} &=& {\rm softmax}(\mathbf{S}\cdot\mathbf{S}^{\top})\cdot\mathbf{S}\\
    \mathbf{S}^{''} &=& {\rm softmax}(\mathbf{S}\cdot\mathbf{P}^{\top})\cdot\mathbf{P}\\
    \mathbf{P}^{'} &=& {\rm softmax}(\mathbf{P}\cdot\mathbf{P}^{\top})\cdot\mathbf{P}\\
    \mathbf{P}^{''} &=& {\rm softmax}(\mathbf{P}\cdot\mathbf{S}^{\top})\cdot\mathbf{S}\\
    \mathcal{L}_{\rm AM3} &=& {\rm MSE}(\mathbf{S}^{'},\mathbf{S}^{''})+{\rm MSE}(\mathbf{P}^{'},\mathbf{P}^{''})
\end{eqnarray}
where ${\rm MSE}$ denotes Mean Squared Error, which requires inputs to have the same length. AM3 efficiently utilises the attention mechanism to avoid relying on alignment information 
for length matching (i.e. in a flat-start manner).

In addition, to further enhance text representation learning, a CTC loss $\mathcal{L}^{\rm Paired}_{\text{CTC}}$ is designed to be computed
based on $\mathbf{P}$, which enforces the consistency between $\mathbf{P}$ and $\mathbf{S}$. 
Neither method includes sequential neural operations and therefore doesn't harm GPU parallel training efficiency.

\vspace{-0.15cm}
\subsection{Training Objective}
\vspace{-0.1cm}
CTC models with FastInject are trained on the paired speech-text data and unpaired text data. On the paired data, a CTC loss, denoted as $\mathcal{L}^{\rm Main}_{\rm CTC}$, is computed on the speech-text pairs as usual. In addition, the
$\mathcal{L}_{\rm AM3}$ and $\mathcal{L}^{\rm Paired}_{\rm CTC}$, as mentioned above,
are also calculated to bridge the modality gap.

On the unpaired text data, the unpaired text representations $\mathbf{U}$ are fed into the CTC-based model as shown in Fig.~\ref{fig:arch} and trained under the CTC supervision $\mathcal{L}^{\rm Unpaired}_{\rm CTC}$. Therefore, the overall training objective can be expressed as:
\begin{equation}
    \mathcal{L} = \mathcal{L}^{\rm Main}_{\rm CTC}+\alpha\cdot(\mathcal{L}^{\rm Paired}_{\rm CTC}+\mathcal{L}^{\rm Unpaired}_{\rm CTC})+\mathcal{L}_{\rm AM3} \label{object}
\end{equation}
where $\alpha$ is a hyper-parameter. 
The three CTC losses share the same linear classifier applied after the acoustic encoder.
\vspace{-0.1cm}
\section{Experiments}
\label{sec:typestyle}
\vspace{-0.1cm}
\subsection{Corpus}
\vspace{-0.05cm}
The paired speech-text data used the “train-clean-100” subset of LibriSpeech \cite{7178964}, a read audiobook corpus, and its dev/test sets were for intra-domain evaluation.
The unpaired text data was taken from GigaSpeech \cite{chen21o_interspeech} training data transcripts (containing audiobook\footnote{Note LibriSpeech evaluation sets are not presented in GigaSpeech \cite{chen21o_interspeech}.}, podcast, and YouTube domains), Fisher training transcripts (telephone domain), and TED-LIUM 2 \cite{rousseau-etal-2014-enhancing} LM text (lecture domain\footnote{The lecture-domain text was sorted by length and one-tenth of the sorted text from the middle was taken so that each domain data is roughly balanced.}). 
This multi-domain unpaired text data was more than 300 times larger than the paired training data transcripts. The test/dev sets of TED-LIUM 2 (lecture), AESRC2020 \cite{9413386} (human-computer interaction), and GigaSpeech (podcast and YouTube) corpora were used for out-of-domain evaluation.
\vspace{-0.1cm}
\subsection{Model Descriptions}
\label{setup}
\vspace{-0.05cm}
Models were built using the ESPnet \cite{Watanabe2018ESPnet} toolkit. Experiments used
1000 modelling units, including 997 BPE units and 3 non-verbal symbols: blank, unknown-character and sos/eos.

A standard CTC model was built with a
12-layer Transformer acoustic encoder (128 attention dimension, 2048 feed-forward dimension, and 4 heads), in which convolutional layers were included for down-sampling by a factor of $1/2$. The S3PRL toolkit \cite{DBLP:conf/interspeech/YangCCLLLLSCLHT21} was used to extract speech features from a fixed WavLM Large model \cite{Chen2021WavLMLS}. A fully-connected layer was added to map the speech features from 1024 to 128 dimensions before being fed into the acoustic encoder.

For the CTC model with FastInject, a 128-dimensional embedding layer and a text encoder, which was the same as the Transformer acoustic encoder but with only 6 layers, were constructed to extract text representations during training and discarded during inference.
Before generating the representations from text, following \cite{zhang2022speechlm, baevski2021unsupervised}, the text was converted into phone units according to a lexicon provided by LibriSpeech and inserting the silence symbol with 25\% probability. When up-sampling the phone units, random repetition was applied using a Gaussian distribution, whose mean and variance values follow \cite{zhang2022speechlm}. The value of $\alpha$ in Eq.~\ref{object} was set to 0.5. 

During training, the ASR models were trained for 25 epochs and
SpecAugment \cite{DBLP:conf/interspeech/ParkCZCZCL19} was used.
Model
parameters from the best 5 epochs were averaged to avoid over-fitting. Greedy search was used for the non-autoregressive CTC model during inference.
Shallow fusion \cite{chorowski2015attention} was also implemented to compare FastInject with the use of an external LM. In this case, a 6-layer Transformer LM was trained on the multi-domain text data for 30 epochs and beam search was used during decoding, where the beam size was 10 and the weight of LM was 0.3.

\vspace{-0.10cm}
\subsection{Experimental Results}
\vspace{-0.1cm}
Experiments compared FastInject with standard CTC training for both intra-domain and out-of-domain scenarios. Ablation studies were conducted to verify the effectiveness of the proposed AM3 and the robustness of FastInject.
\vspace{-0.17cm}
\subsubsection{Intra-domain ASR}
\vspace{-0.1cm}
\begin{table}[t]
    \vspace{-0.25cm}
    \caption{Intra-domain WER on LibriSpeech dev/test sets for CTC models trained on LibriSpeech-100h without LM.}
  \label{tab:ls100-ctc}
  \centering
  \setlength{\tabcolsep}{2.0mm}
  \renewcommand\arraystretch{1.1}
  \begin{tabular}{l | c |c| c| c}
    \Xhline{3\arrayrulewidth}
     \multirow{2}{*}{CTC Models w/o LM}&\multicolumn{2}{c|}{Test}&\multicolumn{2}{c}{Dev}\\
     \cline{2-5}
     &{clean}&{other}&{clean}&{other} \\
     \hline
     wav2vec 2.0 Base \cite{NEURIPS2020_92d1e1eb}&6.1 &13.3& 6.1&13.5\\
     wav2vec 2.0 Large \cite{NEURIPS2020_92d1e1eb}&4.7&9.0&4.6& 9.3\\
     WavLM Base \cite{Chen2021WavLMLS}&5.7& 12.0&--&--\\
     WavLM Base+ \cite{Chen2021WavLMLS}&4.6& 10.1&--&--\\
     token2vec \cite{yue2023token2vec}&5.1&11.8&--&--\\
     \hline
     Standard CTC Model&5.1&9.3&5.0&8.9\\
    CTC with FastInject&\textbf{4.1}&\textbf{7.7}&\textbf{3.9}&\textbf{7.5}\\
    \Xhline{3\arrayrulewidth}
  \end{tabular}
  \begin{tablenotes}
  \footnotesize
  \item{\hspace{-3.5mm}*}{Note the CTC model built in this paper used features from a fixed WavLM Large model. The trainable parameters of the CTC model were 8.53M.}
  \end{tablenotes}
  \vspace{-0.3cm}
\end{table}

\begin{table}[t]
\caption{Out-of-domain WER results on GigaSpeech (Giga),
TED-LIUM 2 (Ted2) and AESRC2020 (AESRC) for CTC models trained from LibriSpeech-100h (LS100) without LM.}
  \label{tab:ls100cross-ctc}
  \centering
  \setlength{\tabcolsep}{0.2mm}
  \renewcommand\arraystretch{1.15}
  \begin{tabular}{l | c c| c c |c c}
    \Xhline{3\arrayrulewidth}
     \multirow{2}{*}{CTC Models}&\multicolumn{2}{c|}{LS100$\Rightarrow$Giga}&\multicolumn{2}{c|}{LS100$\Rightarrow$Ted2}&\multicolumn{2}{c}{LS100$\Rightarrow$AESRC}\\
     &{~\,}{Test}&{~\;}{Dev}&{~\,}{Test}&{~\;}{Dev}&{~\,}{Test}&{~\;}{Dev} \\
    \hline
    Standard CTC&{~\,}21.3&{~\;}22.1&{~\,}11.8&{~\;}12.7&{~\,}16.9&{~\;}15.7\\
    FastInject&{~\,}\textbf{19.0}&{~\;}\textbf{19.5}&{~\,}\textbf{10.0}&{~\;}\textbf{10.1}&{~\,}\textbf{15.0}&{~\;}\textbf{13.9}\\
    \Xhline{3\arrayrulewidth}
  \end{tabular}
  \vspace{-0.3cm}
\end{table}

Table~\ref{tab:ls100-ctc} lists intra-domain ASR results, in which our CTC models achieved good results on the LibriSpeech-100h benchmark compared to recent results\footnote{
Note Table~\ref{tab:ls100-ctc} does not intend to show improving on the previous results, because they focus on the pre-training stage and are not strictly comparable.}. In addition, compared to the standard CTC training, 
the CTC model with FastInject achieved 22\% relative WER reduction (WERR). Therefore, while maintaining a flat-start and efficient ASR training as well as a simple CTC structure, the proposed FastInject approach still boosted CTC performance.

\vspace{-0.17cm}
\subsubsection{Out-of-domain ASR}
\vspace{-0.1cm}
Experiments were conducted to evaluate the out-of-domain ASR performance.
For GigaSpeech and TED-LIUM 2 dev/test sets, unpaired text from the same domain was injected into CTC-based ASR training, but not for AESRC2020.
As shown in Table~\ref{tab:ls100cross-ctc}, compared to standard CTC training, FastInject gave
11.8\%, 21.3\%, and 11.5\% relative WERR on GigaSpeech, TED-LIUM 2, and the AESRC2020 dev/test sets, respectively. This shows that injecting the multi-domain unpaired text data into the CTC-based ASR training improves the general performance, rather than overfitting to the source domain.
In addition, the reduced WER on AESRC2020 dev/test sets indicates that FastInject improves not only the performance within domains represented in the injected text data, but also enhances generalisation to unseen domains.

\vspace{-0.15cm}
\subsubsection{Ablation Studies}
\vspace{-0.05cm}
\begin{table}[t]
    \caption{Ablation studies on the modality matching mechanism of FastInject.}
  \label{ablation_am3}
  \centering
  \setlength{\tabcolsep}{1.1mm}
  \renewcommand\arraystretch{1.12}
  \begin{tabular}{l | c c| c |c| c| c}
    \Xhline{3\arrayrulewidth}
     \multirow{2}{*}{CTC Models}&\multirow{2}{*}{$\mathcal{L}_{\rm AM3}$}&\multirow{2}{*}{$\mathcal{L}^{Paired}_{CTC}$}&\multicolumn{2}{c|}{Test}&\multicolumn{2}{c}{Dev}\\
     \cline{4-7}
     &&&{clean}&{other}&{clean}&{other} \\
    \hline
    Standard CTC&\XSolidBrush&\XSolidBrush&5.1&9.3&5.0&8.9\\
    FastInject&\Checkmark&\Checkmark&\textbf{4.1}&\textbf{7.7}&\textbf{3.9}&\textbf{7.5}\\
    FastInject&\XSolidBrush&\Checkmark&4.3&8.0&4.2&7.8\\
    FastInject&\Checkmark&\XSolidBrush&4.2&7.9&4.1&7.7\\
    FastInject&\XSolidBrush&\XSolidBrush&4.8&8.8&4.7&8.4\\
    \Xhline{3\arrayrulewidth}
  \end{tabular}
  \vspace{-0.3cm}
\end{table}

This paper proposes AM3 to efficiently bridge the modal gap
in a flat-start manner and further computes CTC loss $\mathcal{L}^{\rm Paired}_{\rm CTC}$ from paired text representations to enhance the modality match.
Ablation studies were conducted to show their effectiveness.
As shown in Table~\ref{ablation_am3}, injecting unpaired text into CTC-based ASR training without
$\mathcal{L}_{\rm AM3}$ and $\mathcal{L}^{\rm Paired}_{\rm CTC}$ still resulted in slight improvement with about 5\% relative WERR. This indicates that the CTC model implicitly tries to alleviate the modality gap between speech and text. However, due to a very simple structure, the CTC model, unlike the neural transducer-based model in \cite{sainath2023joist}, is not robust enough to handle the inconsistencies between speech and text representations. Hence, bridging the modality gap is important for the CTC model, and Table~\ref{ablation_am3} shows that both AM3 and 
the computation of $\mathcal{L}^{\rm Paired}_{\rm CTC}$ were very effective in this regard, with AM3 being slightly better. Neither of them relies on previously obtained alignments 
nor harms parallel efficiency, and the best results were obtained when using them together.

\begin{table}[t]
    \caption{Ablation studies on the down-sampling of the text encoder. $\mathbf{S}$ and $\mathbf{P}$ respectively denote paired speech features and paired text representations, as defined in Sec.~\ref{sec:am3}.}
  \label{ablation_cnn}
  \centering
  \setlength{\tabcolsep}{0.4mm}
  \renewcommand\arraystretch{1.12}
  \begin{tabular}{l | c| c| c |c| c| c}
    \Xhline{3\arrayrulewidth}
     \multirow{2}{*}{CTC}&Text Encoder&Length&\multicolumn{2}{c|}{Test}&\multicolumn{2}{c}{Dev}\\
     \cline{4-7}
     &Down-sample&Ratio $\mathbf{S}/\mathbf{P}$&{clean}&{other}&{clean}&{other} \\
    \hline
    Standard&\XSolidBrush&\XSolidBrush&5.1&9.3&5.0&8.9\\
    Joint Train&$1/2$&165\%&\textbf{4.1}&\textbf{7.7}&\textbf{3.9}&7.5\\
    Joint Train&$1/4$&330\%&4.2&7.7&4.0&7.5\\
    Joint Train&$1/1$&82.5\%&4.1&7.8&4.0&\textbf{7.4}\\
    \Xhline{3\arrayrulewidth}
  \end{tabular}
  \vspace{-0.3cm}
\end{table}

To further show the robustness of FastInject, 
the length of the text representations was varied to test whether FastInject is still effective.
In the main experiment, the text encoder contained convolutional layers to achieve down-sampling by a factor of $1/2$. In this section, $1/4$ ratio down-sampling and no down-sampling (i.e. $1/1$) were evaluated.
The results are shown in Table~\ref{ablation_cnn}, where the length ratio $\mathbf{S}/\mathbf{P}$ represents the length ratio of speech features $\mathbf{S}$ to text representations $\mathbf{P}$ on paired training data.
Whichever down-sampling was used, the length of $\mathbf{P}$ was distinctly different from that of $\mathbf{S}$, yet all three scenarios yielded consistent improvements, including $1/4$ down-sampling. This shows the robustness of FastInject and indicates that, when injecting unpaired text into ASR training, it's not always necessary to employ heuristic methods \cite{sainath2023joist, Chen2022MAESTROMS, chen2021injecting} to obtain perfect representation length. 

\vspace{-0.15cm}
\subsubsection{Comparison to External LM}
\vspace{-0.05cm}
\begin{table}[t]
    \caption{WER results on intra (LS100) and out-of-domain (Giga, Ted2, and AESRC) dev/test sets for CTC models trained from LibriSpeech-100h with and without LM. Greedy search was employed when LM was not used and beam search was used when LM was applied.}
  \label{ablation_lm}
  \centering
  \setlength{\tabcolsep}{0.41mm}
  \renewcommand\arraystretch{1.11}
  \begin{tabular}{l | c c| c c| c c |c c |c c}
    \Xhline{3\arrayrulewidth}
     \multirow{2}{*}{CTC}&\multicolumn{2}{c|}{LS100 test}&\multicolumn{2}{c|}{LS100 dev}&\multicolumn{2}{c|}{Giga}&\multicolumn{2}{c|}{Ted2}&\multicolumn{2}{c}{AESRC}\\
     \cline{2-11}
     &{clean}&{other}&{clean}&{other}&test&dev&test&dev&test&dev \\
    \hline
    Standard&\small{5.1}&\small{9.3}&\small{5.0}&\small{8.9}&\small{21.3}&\small{22.1}&\small{11.8}&\small{12.7}&\small{16.9}&\small{15.7}\\
    \quad+LM&\small{3.8}&\small{7.3}&\small{3.7}&\small{6.9}&\small{17.7}&\small{18.2}&\small{8.8}&\small{9.7}&\small{13.1}&\small{11.9}\\
    FastInject&\small{4.1}&\small{7.7}&\small{3.9}&\small{7.5}&\small{19.0}&\small{19.5}&\small{10.0}&\small{10.1}&\small{15.0}&\small{13.9}\\
    \quad+LM&\textbf{\small{3.3}}&\textbf{\small{6.4}}&\textbf{\small{3.2}}&\textbf{\small{6.1}}&\textbf{\small{16.2}}&\textbf{\small{16.5}}&\textbf{\small{7.8}}&\textbf{\small{8.0}}&\textbf{\small{12.2}}&\textbf{\small{11.0}}\\
    \Xhline{3\arrayrulewidth}
  \end{tabular}
  \vspace{-0.3cm}
\end{table}

Table~\ref{ablation_lm} compares
FastInject 
with shallow fusion that uses an external LM, as described in Sec.~\ref{setup}.
Intuitively, when CTC uses an external LM and is decoded via beam search, its performance surpasses greedy search-based results with FastInject, though not always significantly according to a statistical test \cite{115546}. However, this turns the entire system into an autoregressive one, losing the advantage of the CTC model fast decoding speed as a non-autoregressive model.
In contrast, FastInject is more flexible because it retains all the properties of the CTC model during decoding. In addition,
FastInject could also use an external LM to further enhance performance and give 17.5\% relative WERR compared to standard CTC.

\vspace{-0.15cm}
\section{Conclusions}
\label{sec:print}
\vspace{-0.05cm}
This paper proposes a flat-start joint training method, named FastInject, to inject unpaired text data into CTC-based ASR training. 
FastInject inputs text representations into the CTC model along with speech features, before which text units are pre-upsampled.
An attention-based modality matching mechanism (AM3) is proposed to efficiently bridge the gap between speech and text modalities, while maintaining flat-start E2E training and parallel efficiency.
Experiments show that FastInject gave 22.0\% and 20.4\% 
relative WER reductions on intra and out-of-domain test sets, respectively.

\vfill
\pagebreak

\small
\linespread{0.97}\selectfont
\bibliographystyle{ieeetr}
\bibliography{strings,refs}

\end{document}